# Successive spin-flop transitions of Neel-type antiferromagnet Li$_2$MnO$_3$ single crystal with honeycomb-lattice

K. Balamurugan,[1,2] Sang-Hyun Lee,[1,3] Jun-Sung Kim,[4] Jong-Mok Ok,[4] Youn-Jung Jo,[5] Young-Mi Song,[6] Shin-Ae Kim,[7] E. S. Choi,[8] Manh Duc Le,[1,2] and Je-Geun Park[1,2,*]

[1]Center for Correlated Electron Systems, Institute for Basic Science (IBS), Seoul 151-747, Korea.

[2]Department of Physics and Astronomy, Seoul National University, Seoul 151-747, Korea.

[3]Department of Physics, SungKyunKwan University, Suwon 440-746, Korea.

[4]Department of Physics, Pohang University of Science and Technology, Pohang 790-784, Korea.

[5]Department of Physics, Kyungpook National University, Daegu 702-701, Korea.

[6]National Center for Inter-University Research Facilities, Seoul National University, Seoul 151-747, Korea.

[7]Neutron Science Division, Korea Atomic Energy Research Institute, Daejeon 305-353, Korea.

[8]National High Magnetic Field Laboratory, Florida State University, Tallahassee, FL 32310, USA.

**PACS titles:** Spin arrangements in magnetically ordered materials, Exchange and super-exchange interactions, Magnetic phase boundaries including meta-magnetism.

**PACS codes:** 75.25.-j, 75.30.Et, 75.30.Kz

**Abstract**

We have carried out high magnetic field studies of single-crystalline $Li_2MnO_3$, a honeycomb lattice antiferromagnet. Its magnetic phase diagram was mapped out using magnetization measurements at applied fields up to 35 T. Our results show that it undergoes two successive meta-magnetic transitions around 9 T fields applied perpendicular to the *ab*-plane (along the *c\**-axis). These phase transitions are completely absent in the magnetization measured with field applied along the *ab*-plane. In order to understand this magnetic phase diagram, we developed a mean-field model starting from the correct Néel-type magnetic structure, consistent with our single crystal neutron diffraction data at zero field. Our model calculations succeeded in explaining the two meta-magnetic transitions that arise when $Li_2MnO_3$ enters two different spin-flop phases from the zero field Néel phase.

**1. Introduction**

The honeycomb lattice has the smallest number of (three) nearest neighbors that is possible for two dimensional systems. Materials with the honeycomb lattice structure have attracted considerable interest over the years in the condensed matter community, not least for the discovery of massless Dirac fermions in graphene.[1,2] At the same time, honeycomb lattice consisting of magnetic ions are the focus of some interesting ideas such as the Kitaev model,[3] wherein frustrated, directional anisotropic nearest neighbor interactions yield a spin liquid ground state from which exotic quasiparticles called anyons may emerge. Moreover, these may serve as the basis for fault-tolerant quantum computers.[4] Another interesting quantum spin liquid phase was also reported for the Hubbard Model on a honeycomb lattice,[5] whilst a topological insulating state has been discussed in cases where the spin-orbit coupling (SOC) is sufficiently strong, such as $Na_2IrO_3$.[6,7]

Honeycomb lattice compounds containing $Ir^{4+}$ ions (such as $Li_2IrO_3$ and $Na_2IrO_3$), with a large SOC leading to an effective total angular momentum, $J_{eff} = 1/2$,[8-12] have drawn both theoretical and experimental interests in recent years, as they are seen to be probable test beds for the Kitaev model. Nevertheless, these $A_2IrO_3$ compounds (with $A$ = Li and Na) differ from the ideal Kitaev model in that isotropic Heisenberg exchange interactions, arising mainly from direct exchange between nearest neighbor Ir-ions, compete with the directional anisotropic Kitaev interactions. The ground state of such a Heisenberg-Kitaev (HK) model may be one of two spin-liquid or four long-range ordered phases depending on the relative strengths of the Kitaev and the Heisenberg interactions,[13] whilst the $A_2IrO_3$ systems were identified to be long-range ordered zigzag-type antiferromagnets (AFMs).[14,15]

In order to understand better the physics of the honeycomb lattice at a strong SOC limit, it can be a useful and, at the same time, interesting exercise to investigate honeycomb lattice materials at a weak SOC limit. This is an approach we adopted here by examining $Li_2MnO_3$ of such an example thoroughly. $Li_2MnO_3$ also belongs to the "213" honeycomb structure family of compounds[16] with the general formula $A_2TMO_3$.[17] It crystallizes in the monoclinic $C2/m$ space group. Due to its interesting electrochemical activity,[18] $Li_2MnO_3$ has been widely studied for applications in Li-batteries. Below the Néel temperature, $T_N$ = 36.5 K, $Li_2MnO_3$ exhibits an antiferromagnetic ordering of the magnetic moments of $Mn^{4+}$ ions.[19] As shown in Fig. 1(a), it has an alternate stacking of a layer of $Li(1)O_6$ octahedron occupying the center of a honeycomb-like structure formed by $MnO_6$ edge sharing octahedrons and another layer of $Li(2)O_6$ and $Li(3)O_6$ edge sharing octahedrons. The $Mn^{4+}$ ions in $Li_2MnO_3$ occupying the centers of $MnO_6$ octahedrons set up a modified honeycomb lattice network (in the top and bottom *ab*-planes) which fairly mimics the 2D-honeycomb lattice structure of graphene. Therefore, both Mn–O–Mn super-exchange and Mn–Mn Heisenberg direct exchange interactions are simultaneously possible in the *ab*-plane of $Li_2MnO_3$ [see Fig. 1(b)]. Because the $MnO_6$ layers are well separated by the $LiO_6$ octahedral layer, $Li_2MnO_3$ exhibits magnetic properties that are quasi-2D in nature. The antiferromagnetic ordering of $Li_2MnO_3$ single crystal and some of its physical properties have already been reported.[20] In this paper, we present our experimental investigations on the magnetic spin structure and magnetic field (*H*) induced spin-flop (SF) phase transition of $Li_2MnO_3$ single crystal. We construct a full *H-T* phase diagram from the field, temperature (*T*) dependent magnetization and the SF transitions of $Li_2MnO_3$ single crystal measured up to 35 T. The experimental results are compared with the results of mean field model calculations based on a simple Heisenberg Hamiltonian with a single ion anisotropy term.

## 2. Experimental and computational methods

The $Li_2MnO_3$ single crystals were grown by a two-step flux method.[20] In the first step, polycrystalline $Li_2MnO_3$ powder was prepared from $Li_2CO_3$ (99.997 %) and $MnO_2$ (99.99 %). The stoichiometrically mixed starting materials (with 10 mol% of excess $Li_2CO_3$) were pressed into a pellet and placed in an alumina crucible for heat treatments. A high temperature solid state reaction was carried out by heating the pellet to 1027 °C at the rate of 100 °C/h and dwelling for 48 h. Thereafter the polycrystalline $Li_2MnO_3$ sample was cooled to room temperature at the rate of 60 °C/h. The sample was examined for phase purity and crystallinity using a table-top X-ray diffractometer (Rigaku Miniflex II). In the second step, polycrystalline $Li_2MnO_3$ powder was mixed in a plastic bottle with the flux ($Li_2CO_3$ premixed with finely ground $B_2O_3$) in the molar ratio of 1:(2.76:2.39) respectively. This mixture was transferred to a platinum (Pt) crucible and closed using a suitable Pt-lid. The $Li_2MnO_3$ single crystals were grown in the Pt-crucible using the following heating profile: (i) heating from room temperature (RT) to 1092 °C at the rate of 100 °C/h, (ii) dwelling at 1092 °C for 10 h, (iii) cooling to 720 °C at the rate of 2 °C/h, (iv) dwelling at 720 °C for a short period of 10 minutes and (v) natural cooling to RT (by turning the furnace off).

Single crystal neutron diffraction (ND) data of $Li_2MnO_3$ at 10 K was collected using a four circle diffractometer (FCD) having Ge (3 1 1) monochromator and with neutron beam of wavelength, $\lambda_n$ = 1.8343 Å (HANARO reactor, Korea). The ND data was fitted using the FullProf program for different probable magnetic structure models.[21,22] High field magnetization at different temperatures was measured using the National High Magnetic Field Laboratory (NHMFL), Florida (USA), with applied fields up to 35 T. Additional detailed temperature-dependent magnetization measurements were carried out up to 14 T using the vibrating sample magnetometer (VSM) option of a physical property measurement system (PPMS), Quantum Design. We also performed mean-field model calculations for the magnetization of $Li_2MnO_3$ single crystal using the *McPhase* software suite.[23]

## 3. Results
### 3.1 Magnetic structure of $Li_2MnO_3$

As stated in the introduction, the antiferromagnetic structure of $Li_2MnO_3$ has already been reported by Lee *et al*.[20] However, following a recent report[24] that presents a contradicting $C_x$-type AFM model structure that is claimed to fit the muon-spin rotation and relaxation ($\mu^+$SR) experimental data, we were prompted to reinvestigate the magnetic structure of $Li_2MnO_3$ using the single crystal ND data: we measured 320 magnetic Bragg peaks and used 221 independent peaks in our analysis. We attempted to fit the experimental ND data (collected at 10 K) using the following magnetic model structures: $F_x$, $F_y$, $F_{xz}$ ($F_z$), $C_x$, $C_y$, and $C_{xz}$ ($C_z$) -type AFMs. Here, the symbols $F$ and $C$ denote, respectively, ferromagnetic and antiferromagnetic arrangements (interalayer coupling) of magnetic moments in the $ab$-plane with AFM interlayer coupling between any two successive $ab$-planes along the $c$-axis. The direction of the magnetic moments are indicated by the symbols $x$, $y$, $z$ and $xz$ corresponding to the crystallographic $a$-, $b$-, $c$- axes and the $ac$-plane of $Li_2MnO_3$ respectively. The $F_y$, $F_{xz}$, $C_y$ and $C_{xz}$ magnetic model structures have, respectively, the $\Gamma_{1g}$, $\Gamma_{3g}$, $\Gamma_{4u}$ and $\Gamma_{2u}$ symmetry.[20] With the $C_{xz}$-type AFM spin structure, we obtained a best agreement between calculated structure factor ($F^2_{calc}$) and observed structure factor ($F^2_{obs}$) of $Li_2MnO_3$. A schematic drawing of the $C_{xz}$-type AFM spin structure is shown in Fig 2 together with our fitting results for all other models for references. Thus, we again confirm that the magnetic structure of $Li_2MnO_3$ is $C_{xz}$-type AFM with the $\Gamma_{2u}$ symmetry.

In $Li_2MnO_3$, in addition to the nearest neighbor AFM coupling in the $ab$-plane, the magnetic moments have antiferromagnetic (interlayer) coupling along the $c$-axis which doubles the $c$-axis length of the magnetic unit cell with respect to that of the crystallographic (or chemical) unit cell. The refined magnetic moment per $Mn^{4+}$ ion [$\mu_{ord} \approx 2.29(1)$ $\mu_B$] has 0.67(3), 0 and 2.43(1) $\mu_B$ as the components along the $a$-, $b$- and $c$- axes of the unit cell, respectively. Since the magnetic structure requires a strong nearest neighbor ($J_1$) antiferromagnetic interaction and the nearest neighbor Mn–Mn distance is close (~2.8 Å), it is likely that the direct exchange interaction dominates over the probable Mn–O–Mn superexchange interaction which may be weak and ferromagnetic, due to the Mn–O–Mn bond angle (~96°) being close to 90°.[25]

**3.2 High field magnetization of Li$_2$MnO$_3$**

The magnetization of Li$_2$MnO$_3$ single crystal was measured at different temperatures with external magnetic fields applied perpendicular and parallel to the *ab*-plane i.e., parallel and perpendicular to the reciprocal lattice vector, *c**. Figure 3(a) shows the variation of *M* with $H \perp$ *ab*-plane (*H* ∥ *c**-axis). At low temperatures ($T < T_N$), there are two field-induced magnetic phase transitions where the magnetization shows a sharp, nonlinear increase with increasing *H*. No such features were observed with *H* ∥ *ab*-plane ($H \perp c^*$) at any temperature, as shown in Fig. 3(b), indicative of spin flop transitions.[26,27] Accordingly we call the corresponding transition fields spin flop fields, $H_{SF1}$ and $H_{SF2}$ respectively. As seen from Fig. 3(a), the values of $H_{SF1}$ and $H_{SF2}$ increase with *T* up to $T_N$. Figure 3(c) shows an expanded view of *M* measured at different temperatures for both increasing and decreasing $H \perp$ *ab*-plane (*H* ∥ *c**), illustrating the hysteretic behavior of the SF transitions between $H_{SF1}$ and $H_{SF2}$. Interestingly, *M* measured at 2, 5 and 10 K show two clear hysteresis loops, with one that is closer to the $H_{SF2}$ enclosing a relatively lower area than the other that is closer to $H_{SF1}$. The hysteresis loops in the *M* vs *H* data indicate first order phase transitions.[25]

Figure 4 shows the temperature dependent magnetization of Li$_2$MnO$_3$ single crystal measured using field-cooled warming protocol with various constant magnetic fields applied perpendicular and parallel to the *ab*-plane (i.e., parallel and perpendicular to *c**-axis). The *M* vs *T* curve measured at *H* = 0.03 T shows a broad maximum around 50 K and a kink at $T_N \approx 36.5$ K, which is more clearly seen for $H \parallel c^*$ than for $H \perp c^*$. As the applied external magnetic field is increased, $T_N$ gradually decreases. Below $T_N$, a minimum is observed in the magnetization for $H \parallel c^* > 8.5$ T and it shifts to higher temperature with increasing field until $H > 11$ T, where no clear (or sharp) minimum can be seen. At higher fields, for example $H = 14$ T, the changes in magnetization corresponding to both $H \parallel c^*$ and $H \perp c^*$ appear to be very similar. The minima in $M$ are due to the SF transitions corresponding to the specific values of $H \perp ab$-plane (i.e., $H \parallel c^*$), because these occur only for $H$ applied (nearly) parallel / antiparallel to the direction of the magnetic moments ($\mu$) of an antiferromagnet with low anisotropy[27] and not for $H \perp \mu$. We note here that, in principle, the SF transitions can be either first order or second order.[28]

### 3.3 Magnetic phase diagram of $Li_2MnO_3$

As shown in Fig. 5, we construct a magnetic phase diagram based on the experimental results: the variation of $T_N$ with $H$ (●-symbol), and the SF transition fields ($H_{SF1}$ and $H_{SF2}$) obtained from $M$ vs $H$, hysteresis measurements (▲- and ▼- symbols) and the $M$ vs $T$ curves at different fixed $H \parallel c^*$ (■- and ★-symbols). The transition from paramagnetic (PM) phase to the antiferromagnetic (AFM) phase is a second-order phase transition whose transition temperature ($T_N$) decreases with increasing $H$. For $T \ll T_N$, the AFM phase is separated from the SF phase by two first-order phase boundary lines (indicated by ▲-, ▼-, ■- and ★- symbols). The lines connecting those data points are guides to the eyes. The inset of Fig. 5 (an expanded view of the phase diagram) shows the trend of merging first-order $H_{SF1}$ and $H_{SF2}$ boundary lines and joining the second-order PM-AFM/SF phase boundary line. The merging point [$(10 < H_t < 11$ T$), T \approx T_N(H_t)$] is a tri-critical point in the $H$-$T$ diagram of $Li_2MnO_3$ which connects the PM, AFM and SF phases. Whilst the high field SF phase should have magnetic moments in the $ab$-plane, the structure of the intermediate phase (at $H_{SF1} < H < H_{SF2}$) was unclear, leading us to perform a mean field analysis.

### 3.4 Mean field analysis and the spin-flop transition of Li$_2$MnO$_3$

The essential features of the observed SF transitions and some other physical properties of Li$_2$MnO$_3$ may be described by a spin, $S = 3/2$ Heisenberg model with weak easy axis anisotropy (along $c^*$-axis), using the following Hamiltonian:

$$H = -\frac{1}{2}\left[ J_1 \sum_{n.} \mathbf{S}_i \cdot \mathbf{S}_j + J_2 \sum_{n.n.} \mathbf{S}_i \cdot \mathbf{S}_j + J_3 \sum_{n.n.n.} \mathbf{S}_i \cdot \mathbf{S}_j + J_c \sum_{n.c.} \mathbf{S}_i \cdot \mathbf{S}_j \right] - K \sum_i (S_i^z)^2 \quad (1),$$

where the summations run over nearest- (n.), next-nearest (n.n.), and next-next-nearest (n.n.n.) neighbor ($i^{th}$ and $j^{th}$) Mn-ions in the $ab$-plane, and nearest neighbors along the $c$-direction (n.c.) with associated exchange parameters $J_1$, $J_2$, $J_3$ and $J_c$ respectively (as shown in Fig. 1b), and $K$ is the single ion anisotropy parameter. As a first step, a single spin-flop transition can be produced by a simplified model with only non-zero $J_1$ and $K$, which may be uniquely defined by the Néel temperature and the critical field value. But, the observed two spin flop transitions require an additional, small, non-zero $J_c$. While the interlayer interaction ($J_c$) is necessary for the $C_{xz}$-type AFM spin structure it also stabilizes an intermediate field structure, wherein only moments on the alternate $ab$-planes have flopped, as illustrated in Fig. 5. The magnitude of $J_c$ determines the difference between the spin-flop fields ($H_{SF2} - H_{SF1}$), and thus may be fixed by the experimental results. Since the next- and next-next -nearest neighbor interalayer interactions ($J_2$ and $J_3$) and the interlayer interaction ($J_c$) likely share similar Mn–O–Li–O–Mn super-exchange pathways that are separated by almost the same distances, they should, a priori, be of similar magnitude. Therefore, we have fixed the values of $J_2$ and $J_3$ at twice the magnitude of $J_c$. Unfortunately this produces a large change in the magnetization at the spin flop transition that is calculated to be approximately twice the measured values. In order to reduce this and to match with the experimentally observed change in the magnetization, we need larger unphysical values of $J_2$ or $J_3$.

Given the above constraints, we found that the following set of exchange and anisotropy constants explains our data better: $J_1 = -0.84$ meV, $J_2 = J_3 = -0.02$ meV, $J_c = -0.01$ meV, and $K = -0.067$ meV. The calculated magnetization is shown as dashed lines (MFCs) in Fig. 3(a,b,c) and Fig. 4, and the calculated magnetic phase diagram is given by the background in Fig. 5. The spin-flop transitions are calculated to be first order at low temperatures, which is consistent with the experimental observations, and apparently merging at $T \approx 16$ K. Above this temperature, as a function of increasing field, the moments rotate smoothly from being perpendicular to the *ab*-plane to parallel to the *ab*-plane, reminiscent of a liquid-gas critical point. However, although the agreement between the theoretical calculations and the experimental results are reasonably good, there are clear disagreements too. A most noticeable case is that the calculated magnetization is bigger than the experimental field dependence of the magnetization data shown in Fig. 3. We have tested several alternative models by varying values of *J* and *K* to resolve this discrepancy before coming to a conclusion that the magnetic moment of Mn ions may as well be effectively smaller in the real material than the spin-only ionic value (3 $\mu_B$). In fact, the ordered moment determined by the ND refinements is not 3 $\mu_B$ but ~2.3 $\mu_B$. This is strong evidence that supports our idea. Moreover, short-ranged fluctuations, often present in two-dimensional spin systems, can reduce the effective spin value too, which, in principle, cannot be accounted in mean-field calculations. Therefore, we repeated the calculations using the following set of parameters for a spin, $S = 1$ model: $J_1 = -1.55$ meV, $J_2 = J_3 = -0.02$ meV, $J_c = -0.025$ meV, and $K = -0.109$ meV. The results of the MFCs using $S = 1$ are shown by the dashed-symbol lines in Fig. 3 which exhibits significant improvement over that of MFCs using $S = 3/2$. (See also calculated $M(T)$ results in Fig. 4). However, we note that the overall phase diagram remains almost unchanged when compared to that for the $S = 3/2$ case. The magnetic structures presented in Fig. 5 are the results of our mean-field calculations, and that corresponding to the low-field region in the H-T phase diagram is indeed consistent with our analysis of the experimental data as shown in Figs. 1 and 2.

## 4. Discussions

It is well known that the SF occurs if an external magnetic field of sufficient strength is applied parallel to one of the two sub-lattice magnetic moments of an antiferromagnet with a small anisotropy of easy axis of magnetization.[26,27] The whole process of SF is an act of reducing the energy of the system which, otherwise, is higher if one of the AFM sub-lattice's moments point antiparallel to a sufficiently strong $H$. The spin-flop transition in $Li_2MnO_3$ single crystal occurs in two steps. In first step, when an external magnetic field of strength $H \geq H_{SF1}$ ($< H_{SF2}$) is present, *only the spins of the alternate ab-planes flop*, because the spin-flop occurs against the single ion anisotropy (SIA) term ($K$) which adds additional energy if the spins in all the *ab*-planes flop for the same strength of $H$. Eventually, the interlayer antiferromagnetic exchange interaction ($J_c$) is weakened in the presence of a strong magnetic field. In second step, when $H$ is increased further to $H \geq H_{SF2}$ ($< H_{FM}$), *the spins of the other alternate ab-planes also flop* because this reduces the total energy of the system (even though it would add a little energy due to $K$). Here, $H_{FM}$ is a possible (hypothetical) field of unknown strength, such that for $H \geq H_{FM} \gg H_{SF2}$, all the spins in all the *ab*-planes *flip* to align parallel to $H$ and establish a field-induced ferromagnetic (FM) phase. Inspecting the high field magnetization data shown in Fig. 3a, we estimate $H_{FM}$ to be in the range of 70 – 100 T.

In fact, it is interesting to note that a FM spin-structure in a honeycomb lattice system (even though a classical ground state) is another possible magnetic phase of the HK model[13,29] that is widely studied at present. It may be possible to observe a field-induced FM phase in a honeycomb lattice system (e.g., Li$_2$MnO$_3$ or a similar one) at extremely high magnetic fields. We observe that the search for honeycomb lattice systems with such field-induced FM phase or FM ground state may lead to the emergence of new applications of magnetism in honeycomb lattice. Perhaps, this is achievable in some new materials with similar structure or materials that offer honeycomb lattice for magnetic ions and have strong single ion anisotropy. A *modified Kitaev Hamiltonian*, introduced by Baskaran *et al.*, has been shown exactly solvable for all half-odd-integer spins and commented as "*it is equivalent to an exponentially large number of copies of spin-1/2 Kitaev Hamiltonians*".[30] This modified spin-$S$ Kitaev Hamiltonian may be thought to replace the original Kitaev interaction terms of the currently considered HK model, so that the modified HK model would be more general and may be applicable to any real honeycomb lattice materials with high spins ($S > 1/2$) also, such as Li$_2$MnO$_3$.

By our high field magnetization studies of Li$_2$MnO$_3$ single crystal and the mean field model calculations, we found that the nearest neighbor interlayer interaction ($J_c$) along the *c*-axis is exhibiting a distinguishable feature in the magnetization and SF transition. This finding is important since the inter-layer coupling is mostly neglected in theoretical studies of the magnetic phase diagram of honeycomb lattice systems. Overall, what we found is that it is essential to comprehend the field-induced phase, which, we think, may well be relevant for other magnetic honeycomb lattice systems. Further, besides the fact that the spins of $Mn^{4+}$ ions in $Li_2MnO_3$ can be treated effectively as classical spins, its Néel-type AFM spin-structure may be thought of one of the classical ground states of HK model with Kitaev interaction of negligible or effectively zero strength relative to the strength of isotropic Heisenberg interaction. This is justifiable because in the HK model reported by Chaloupka *et al*., the Néel-type AFM phase exists for a wide range of "$\varphi$" ($-34° < \varphi < 88°$), where $\varphi$ is a phase angle determined by the relative strength of the Kitaev term ($2K$) with respect to the Heisenberg term ($J$): $\varphi = \cos^{-1}\left(\frac{J}{\sqrt{K^2+J^2}}\right)$.[13] Our analysis discussed so far suggests that the Kitaev term should be very small for $Li_2MnO_3$ compared with the Heisenberg term. Thus, the phase angle ($\varphi$) is almost zero for $Li_2MnO_3$, which produces the Néel phase according to the theoretical results in Refs. 13 and 29: which is in good agreement with our experimental results. What is particularly interesting about our data regarding the general phase diagram of the HK model is the newly discovered field-induced spin-flop phases and the ferromagnetic (FM) phase. First of all, there is no theoretical prediction available at the moment for the HK Hamiltonian plus a Zeeman term so we cannot make a direct comparison with our data. However, it is intriguing that in all theoretical phase diagrams of Refs. 13 and 29 the magnetic structure we found for the spin-flop phase is not found to be stable. Second, our high field data demonstrate that one can adiabatically move from the Néel phase to the FM phase. Therefore, it will be highly interesting to examine the thermodynamics of the HK Hamiltonian plus a Zeeman term as a function of magnetic field. Thus far we believe that our experimental works demonstrated that there is more for future developments, which could offer much better understandings on the magnetism of honeycomb lattice systems with any spin-*S*.

## 5. Summary and conclusion

In summary, we have investigated $Li_2MnO_3$ single crystal as a test bed for the physics of magnetism in honeycomb lattice at a weak SOC limit as it offers a honeycomb lattice for $Mn^{4+}$ ions with spin, $S = 3/2$. $Li_2MnO_3$ has a classical Néel ordered $C_{xz}$-type antiferromagnetic spin structure ($\Gamma_{2u}$ symmetry). The refined resultant magnetic moment per $Mn^{4+}$ ion is $\mu_{ord} \approx 2.29(1)$ $\mu_B$, reduced from the spin-only ionic value (3 $\mu_B$) probably because of the two-dimensional nature of the honeycomb lattice. Further, we have studied high field magnetization of $Li_2MnO_3$ single crystal for magnetic fields up to 35 T applied both parallel and perpendicular to the *ab*-planes. $Li_2MnO_3$ is seen to exhibit two successive magnetic field induced spin-flop phase transitions at $T < T_N$. A magnetic phase diagram of $Li_2MnO_3$ single crystal has been constructed using the high field- and temperature- dependent magnetization data. The spin-flop phase transition and the other magnetization properties of $Li_2MnO_3$ single crystal can be described well by a honeycomb lattice system of an *effective* spin, $S = 1$ model based on a simple Heisenberg exchange interaction Hamiltonian with a single ion anisotropy term. In the mean field analysis, the spin-flop transition has been found to occur in two successive first-order phase transitions at lower temperatures; surprisingly, the inter-layer coupling is seen to play an essential role for the spin-flop transition. This is seen to have good agreement with the two hysteresis loops observed in our $M$ vs $H$ data.

**Acknowledgements**

We acknowledge Jiyeon Kim, Choongjae Won, and Namjung Hur for their contribution at the beginning of the project, and Gun Sang Jeon for helpful discussion. This work was supported by the Institute for Basic Science (IBS) in Korea.

# References

*To whom all correspondence should be made: jgpark10@snu.ac.kr

[1] K. S. Novoselov, A. K. Geim, S. V. Morozov, D. Jiang, M. I. Katsnelson, I. V. Grigorieva, S. V. Dubonos, and A. A. Firsov, Nature **438**, 197 (2005).

[2] Y. Zhang, Y.-W. Tan, H. L. Stormer, and Philip Kim, Nature **438,** 201 (2005).

[3] A. Kitaev, Ann. Phys. **321**, 2 (2006).

[4] A.Y. Kitaev, Ann. Phys. **303**, 2 (2003).

[5] Z. Y. Meng, T. C. Lang, S. Wessel, F. F. Assaad, and A. Muramatsu, Nature **464**, 847 (2010).

[6] A. Shitade, H. Katsura, J. Kuneš, X.-L. Qi, S.-C. Zhang, and N. Nagaosa, Phy. Rev. Lett. **102**, 256403 (2009).

[7] C. H. Kim, H. S. Kim, H. Jeong, H. Jin, and J. Yu, Phy. Rev. Lett. **108**, 106401 (2012).

[8] B. J. Kim, H. Ohsumi, T. Komesu, S. Sakai, T. Morita, H. Takagi, and T. Arima, Science **323**, 1329 (2009).

[9] J. Chaloupka, G. Jackeli, and G. Khaliullin, Phys. Rev. Lett. **105**, 027204 (2010).

[10] J. Reuther, R. Thomale, and S. Trebst, Phys. Rev. B, **84**, 100406(R) (2011).

[11] Y. Singh, S. Manni, J. Reuther, T. Berlijn, R. Thomale, W. Ku, S. Trebst, and P. Gegenwart, Phys. Rev. Lett. **108**, 127203 (2012).

[12] H. Gretarsson, J. P. Clancy, X. Liu, J. P. Hill, E. Bozin, Y. Singh, S. Manni, P. Gegenwart, J. Kim, A. H. Said, D. Casa, T. Gog, M. H. Upton, H.-S. Kim, J. Yu, Vamshi M. Katukuri, L. Hozoi, J. van den Brink, and Y.-J. Kim, Phys. Rev. Lett. **110**, 076402 (2013).

[13] J. Chaloupka, G. Jackeli, and G. Khaliullin, Phys. Rev. Lett. **110**, 097204 (2013).

[14] X. Liu, T. Berlijn, W.-G. Yin, W. Ku, A. Tsvelik, Y.-J. Kim, H. Gretarsson, Y. Singh, P. Gegenwart, and J. P. Hill, Phys. Rev. B **83**, 220403 (R) (2011).

[15]F. Ye, S. Chi, H. Cao, B. C. Chakoumakos, J. A. Fernandez-Baca, R. Custelcean, T. F. Qi, O. B. Korneta, and G. Cao, Phys. Rev. B **85**, 180403(R) (2012).

[16]I. I. Mazin, S. Manni, K. Foyevtsova, H. O. Jeschke, P. Gegenwart, and R. Valentí, Phys. Rev. B **88**, 035115 (2013).

[17]The *TM*-site is occupied by any 3*d*, 4*d* or 5*d* elements that exhibits stable +4 oxidation states.

[18]D. Y. W. Yu, K. Yanagida, Y. Kato, and H. Nakamura, J. Electrochem. Soc. **156** (6), A417 (2009).

[19]P. Strobel, and B. Lambert-Andron, J. Solid State Chem. **75**, 90 (1988).

[20]S. Lee, S. Choi, J. Kim, H. Sim, C. Won, S. Lee, S. A. Kim, N. Hur, and J.-G. Park, J. Phys.: Condens. Matter **24**, 456004 (2012).

[21]J. Rodríguez-Carvajal, Physica B **192**, 55 (1993).

[22]T. Roisnel, J. Rodriguez-Carvajal, Proc. Eur. Powder Diffr. Conf. **7**, 118 (2000).

[23]M. Rotter, J. Magn. Magn. Mater. **272–276**, e481 (2004).

[24]J. Sugiyama, K. Mukai, H. Nozaki, M. Harada, M. Månsson, K. Kamazawa, D. Andreica, A. Amato, and A. D. Hillier, Phys. Rev. B **87**, 024409 (2013).

[25]P. W. Anderson, Phys. Rev. **79**, 350 (1950).

[26]B. R. Morrison, Phys. Stat. Sol. (B) **69**, 581 (1973).

[27]H. J. M. de Groo and L. J. de Jongh, Physica B **141**, 1 (1986).

[28]I. S. Jacobs and P. E. Lawrence, Phys. Rev. **164**, 866 (1967).

[29]J. G. Rau, E. K-H. Lee, and H-Y. Kee, Phys. Rev. Lett. **112**, 077204 (2014).

[30]G. Baskaran, D. Sen, and R. Shankar, Phys. Rev. B **78**, 115116 (2008).

**Figure captions**

FIG. 1. (Color online) (a) A polyhedral view of $Li_2MnO_3$ single crystal (viewed perpendicular to the *ab*-plane). (b) The $C_{xz}$-type antiferromagnetic spin structure (unit cell) of $Li_2MnO_3$ with only Mn atoms shown for better clarity. The nearest-, next-nearest- and next-next-nearest neighbor exchange interactions ($J_1$, $J_2$ and $J_3$) between Mn atoms in the *ab*-plane are shown by double headed solid, dashed and dotted arrows, respectively. The possible exchange interaction along the *c*-axis ($J_c$) is shown by a double headed dashed-dotted arrow.

FIG. 2. (Color online) Result of magnetic structure refinements using neutron diffraction data of $Li_2MnO_3$ single crystal. The dashed lines in each graph are the reference lines of perfect match between the squared observed ($F^2_{obs}$) and calculated ($F^2_{cal}$) structure factors. Schematic diagrams are given for each magnetic structures used in the refinement.

FIG. 3. (Color online) Variation of high field magnetization of $Li_2MnO_3$ single crystal with magnetic field applied (a) parallel and (b) perpendicular to *c\**-axis measured (only for decreasing field) at different temperatures. (c) An expanded view of high field magnetization measured for both increasing and decreasing field applied parallel to *c\**-axis. The corresponding magnetization calculated using mean field models (MFC) having spin $S = 3/2$ and $S = 1$ are shown by dashed lines and dashed-symbol lines respectively.

FIG. 4. (Color online) Temperature dependence of magnetization of $Li_2MnO_3$ single crystal measured for different values of magnetic field applied both parallel and perpendicular to *c\**-axis. In this picture, unless specified otherwise, the direction of the magnetic field is parallel to *c\**-axis.

The corresponding magnetization calculated using mean field models (MFC) having spin $S = 3/2$ and $S = 1$ are shown by dashed lines and dashed-symbol lines respectively.

FIG. 5. (Color online) Magnetic phase diagram of $Li_2MnO_3$ single crystal. The data points are derived from the experimental magnetization measured at either fixed temperatures (▲ and ▼ symbols) or fields (■ and ★ symbols) while sweeping the other parameter. (The lines connecting the data points are guides to the eyes.) The phase diagram constructed from our mean field model calculations (shaded area) is shown as the background for the phase diagram of experimental data. The magnetic spin structures of the ordered AFM (obtained from the analysis of experimental single crystal neutron diffraction data), spin flop (SF) and intermediate SF phases (obtained in the mean field analysis) are also shown. The inset shows an expanded view of the phase diagrams with a trend of merging first-order $H_{SF1}$ and $H_{SF2}$ boundary lines and joining the second-order PM-AFM/SF phase boundary line.

**Figures**

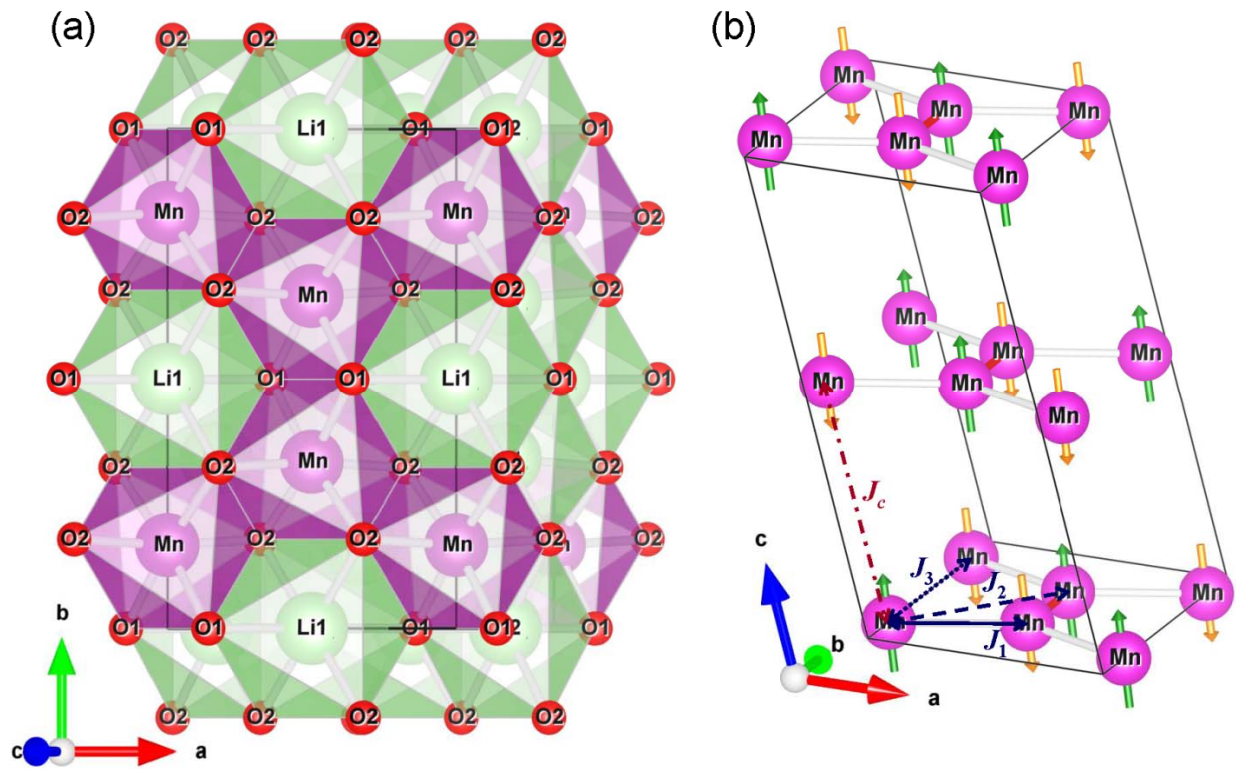

FIG. 1

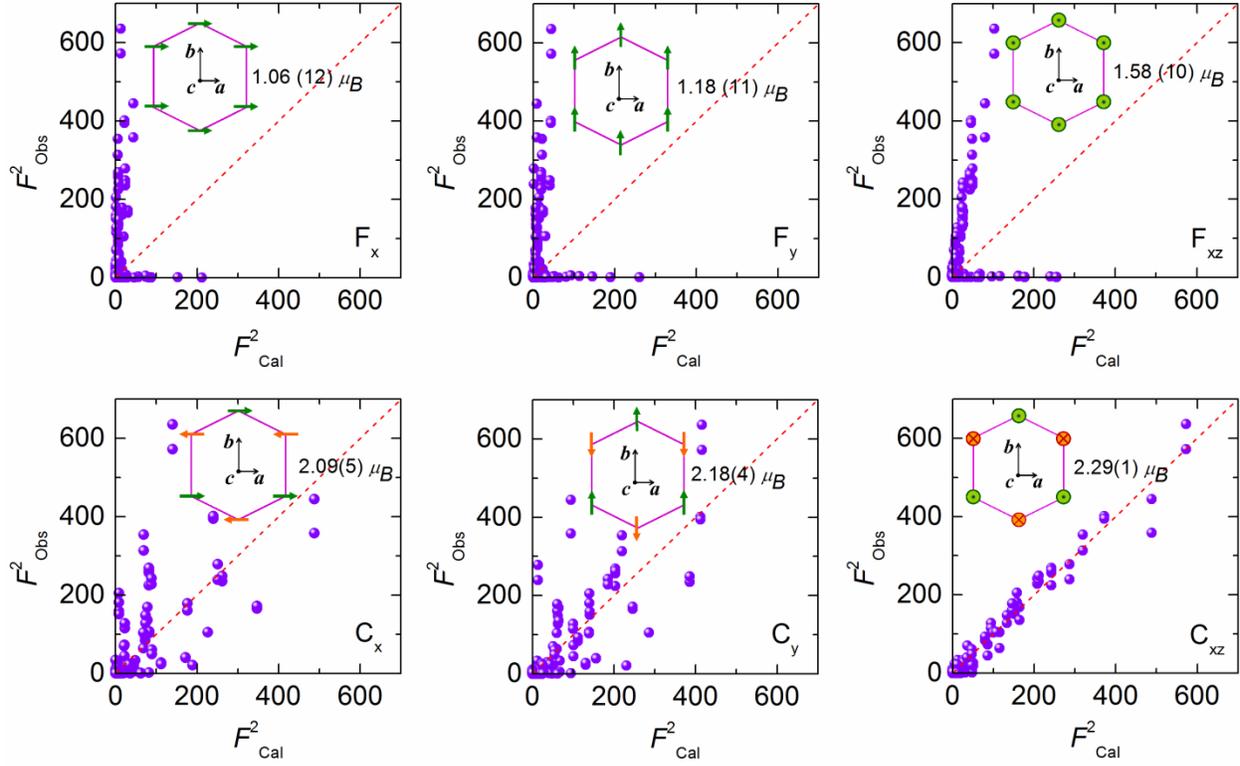

FIG. 2

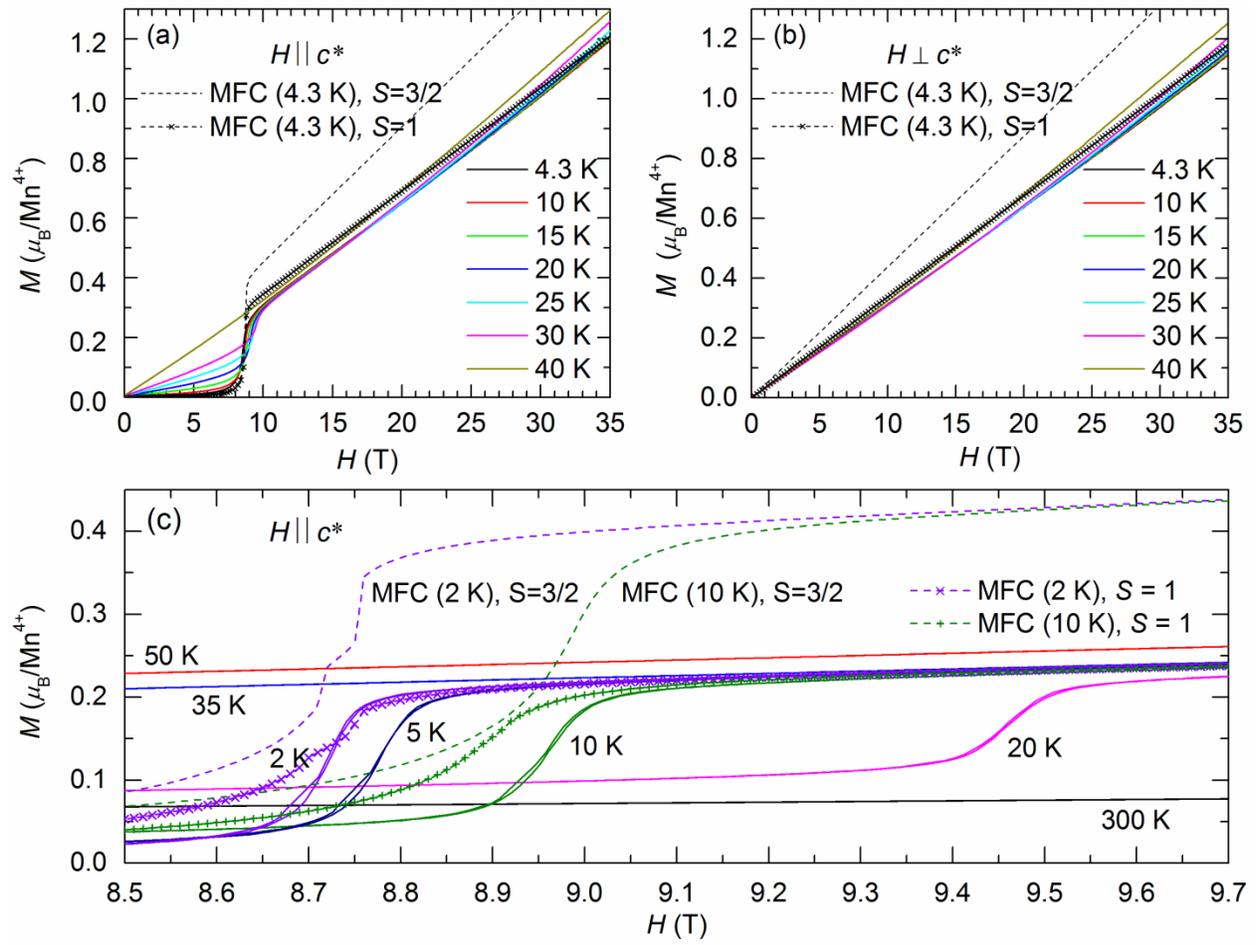

FIG. 3

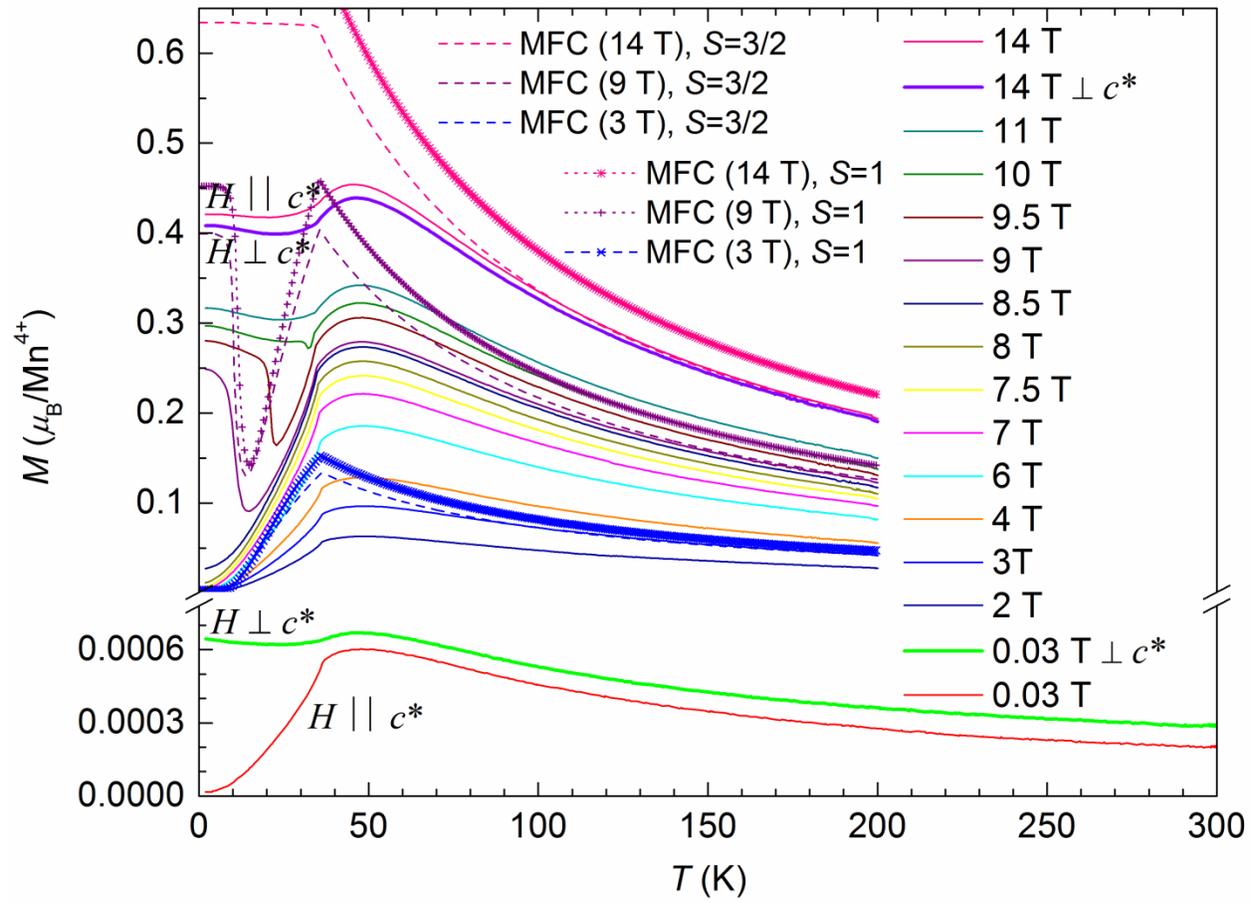

FIG. 4

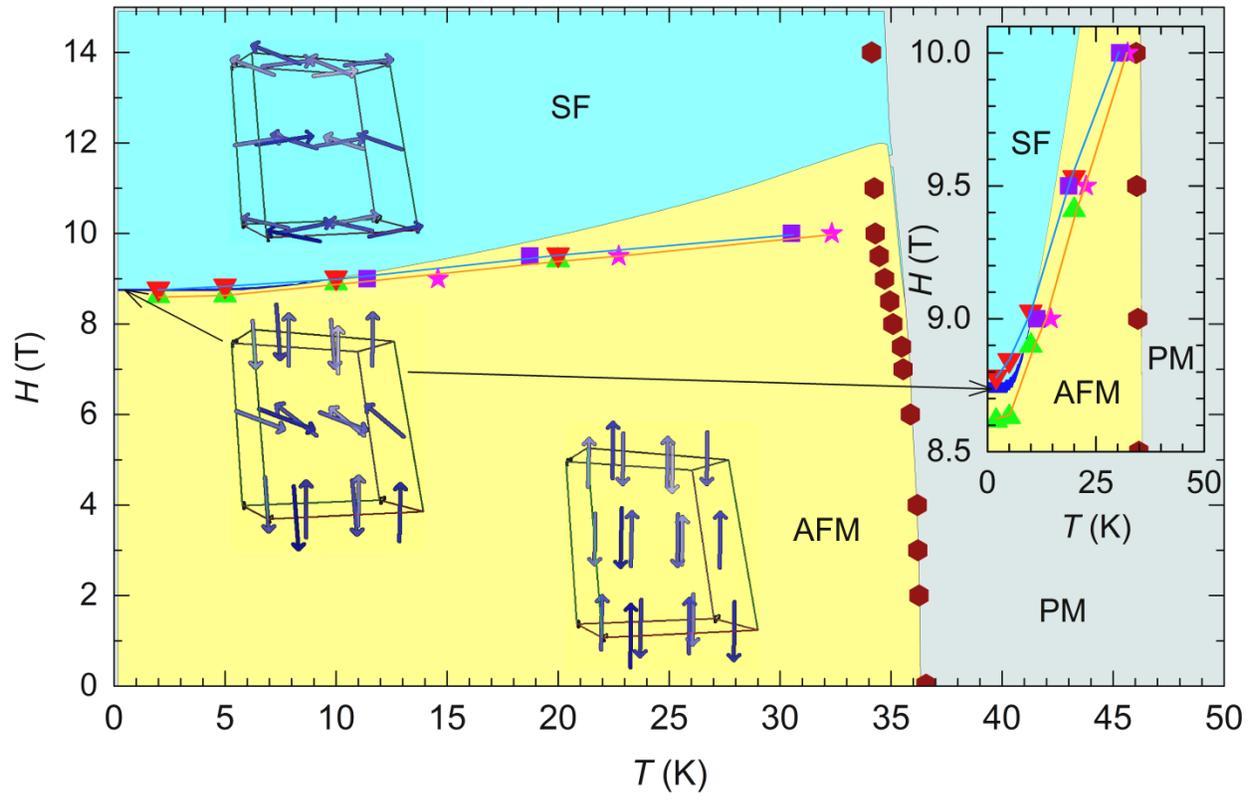

FIG. 5